\documentclass[12pt]{article}

\usepackage{epsfig}

{\end{list}}
\newcounter{enumct}
\newenvironment{Enumerate}{\begin{list}{\arabic{enumct}.}%
{\usecounter{enumct}\setlength{\topsep}{0.2mm}%
\setlength{\partopsep}{0.2mm}\setlength{\itemsep}{0.2mm}%
\setlength{\parsep}{0.2mm}}}{\end{list}}

\newcommand{\captive}[1]{\rule{5mm}{0mm}%
\begin{minipage}{150mm}\caption[small]{#1}\end{minipage}}
 
\begin{document}

\sloppy


\begin{flushright}
LBNL--42524\\
November, 1998
\end{flushright}

\vspace{-6mm}
\begin{center}
{\LARGE\bf Two--Photon Physics in Nucleus-Nucleus}\\[2mm]
{\LARGE\bf Collisions at RHIC}\\[5mm]
{\Large Joakim Nystrand and Spencer Klein} \\[3mm]
{\it Lawrence Berkeley National Laboratory,}\\[1mm]
{\it Berkeley, California 94720, U.S.A.}\\[1mm]
{\it E-mail: JINystrand@lbl.gov, SRKlein@lbl.gov}\\[3mm]

{\large and}\\[3mm]

{\Large the STAR Collaboration
\footnote{The STAR Collaboration list can be found 
(under "organization") at: \newline
http://rsgi01.rhic.bnl.gov/star/starlib/doc/www/star.html}
}\\[3mm]

{\small\it Invited talk presented at Workshop on Photon Interactions and the 
Photon Structure, \newline Lund, Sweden, September 10--13, 1998}\\[4mm]

{\bf Abstract}\\[1mm]
\begin{minipage}[t]{140mm}
Ultra--relativistic heavy--ions carry strong electromagnetic and nuclear
fields. Interactions between these fields in peripheral nucleus--nucleus 
collisions can probe many interesting physics topics. This presentation 
will focus on coherent two--photon and photonuclear processes at RHIC. 
The rates for these interactions will be high. The coherent coupling of all 
the protons in the nucleus enhances the equivalent photon flux  
by a factor $Z^2$ up to an energy of $\sim$~3~GeV. The plans for studying 
coherent interactions with the STAR experiment will be discussed. 
Experimental techniques for separating signal from background will be 
presented.
\end{minipage}\\[5mm]

\rule{160mm}{0.4mm}

\end{center}

\vspace{-7mm}
\section{The Relativistic Heavy-ion Collider (RHIC) and the STAR Experiment}

The Relativistic Heavy-Ion Collider (RHIC), which is now under construction 
at the Brookhaven National Laboratory, is designed to accelerate protons 
and nuclei (A=1--197) to energies of 100 -- 250 A GeV \cite{RHIC}. 
The center-of-mass collision energies will be an order of magnitude 
higher than has previously been available for heavy nuclei. 
Some of the possible beams, with corresponding beam energies and 
luminosities, are listed in Table~1. RHIC will begin 
operation in 1999.

STAR (Solenoidal Tracker at RHIC) is one of two large experiments 
at RHIC\cite{STAR}. STAR is primarily designed to study hadronic
observables over a wide region of phase space. The main detector is a 
large, cylindrical Time Projection Chamber (TPC), which is placed inside a 
0.5~T solenoidal magnet. Tracks of charged particles produced in the 
interactions
will be reconstructed in the TPC, and the particle momenta will be determined
from the curvature in the magnetic field. Information on energy
loss in the TPC gas may be used to identify low--momentum particles.
The main TPC roughly covers the 
pseudorapidity range $|\eta|<2.0$ ($\eta = - \tan ( \theta / 2 )$, where 
$\theta$ is the polar emission angle). 

Other detectors in STAR are the forward TPCs (FTPC), the silicon vertex
tracker (SVT), and the electromagnetic calorimeter (EMC). The forward TPCs 
will complement the main TPC by detecting charged particles in the range  
$2.5 \leq \mid \eta \mid \leq 3.75$. 
A Time--of--Flight (TOF) system and a Ring--Imaging Cherenkov (RICH) 
detector have been proposed and would provide improved particle 
identification at high momenta. Figure~1 shows a picture of STAR.

STAR will have a flexible multi--level trigger. In the lower trigger
levels, the trigger information will be provided by the central trigger 
barrel (CTB) and the TPC anode wires, operating as wire chambers. The CTB
consists of 240 scintillators covering $| \eta | <$~1; the TPC anode wires 
will provide multiplicity information in the range $1<|\eta|<2$. 
In the highest trigger level TPC tracking 
information will also be available. In addition, zero--degree
calorimeters along the beam lines on either side of the experiment (for 
detection of neutrons in nuclear breakup) are used for triggering.

Section 5 will discuss the experimental techniques 
for studying coherent interactions within STAR.
The analyses presented there will be based on a year 1 configuration of STAR
consisting of the main TPC, forward TPCs, CTB, and the TPC anode
wire read--out.
Without the EMC, only final states consisting exclusively of charged particles 
will be considered.

\begin{center}
\begin {table} [t!] \begin{center}
\begin{tabular} {lccccc} \hline
Projectile & Z & A & Kinetic Energy & Lorentz factor, $\gamma$ & Luminosity \\ 
           &   &   & $[$A GeV$]$    &       & $[$cm$^{-2}$ s$^{-1}]$ \\ \hline
p          & 1 & 1 & 251 & 268          & 1.4 $\cdot$ 10$^{31}$ \\
O          & 8 &16 & 125 & 135          & 9.8 $\cdot$ 10$^{28}$ \\
Si         & 14&28 & 125 & 135          & 4.4 $\cdot$ 10$^{28}$ \\
Cu         & 29&63 & 115 & 126          & 9.5 $\cdot$ 10$^{27}$ \\
I          & 53&127& 104 & 113          & 2.7 $\cdot$ 10$^{27}$ \\
Au         & 79&197& 100 & 108          & 2.0 $\cdot$ 10$^{26}$ \\ \hline
\end{tabular}
\label{RHIC}
\captive{Possible beam projectiles and maximum energies and luminosities at
RHIC, from the RHIC conceptual design report\cite{RHIC}.}
\end{center}
\end{table}
\end{center}

\section{Coherent Nuclear Interactions}

Coherent nuclear interactions are defined as reactions between the fields
of the nuclei, whereby the fields couple coherently to all the nucleons. 
The nuclei do not directly participate in these interactions, but act as 
sources of fields. In the reactions, 
\begin{equation}
   A + A \rightarrow A + A + X \; ,
\end{equation}
a final state X is produced, while the nuclei normally remain in 
their ground state. 

These reactions occur through the collision of two exchange particles. 
For purely electromagnetic interactions, the exchange particles are
photons, and the process corresponds to a two-photon interaction. 
For the strong nuclear force, the exchange particles are in principle
gluons. Gluons cannot, however, couple coherently to several nucleons,
since they have a color charge. The colorless exchange particle of the 
nuclear interaction is the Pomeron, which, under certain 
circumstances, might be interpreted as a pair of gluons or a gluon 
ladder\cite{Pomeron}. Colorless nuclear interactions may also be
mediated by mesons. Coherent nuclear interactions can thus be of the following
types: two--photon, photon-Pomeron/meson (photonuclear) and 
Pomeron/meson--Pomeron/meson. The cross section for Pomerom--Pomeron reactions
is expected to be low\cite{Schramm}.
In the rest of this paper, only two-photon 
and photonuclear interactions will be discussed further.

\begin{figure}
\epsfxsize=0.59\textwidth
\centerline{\epsffile{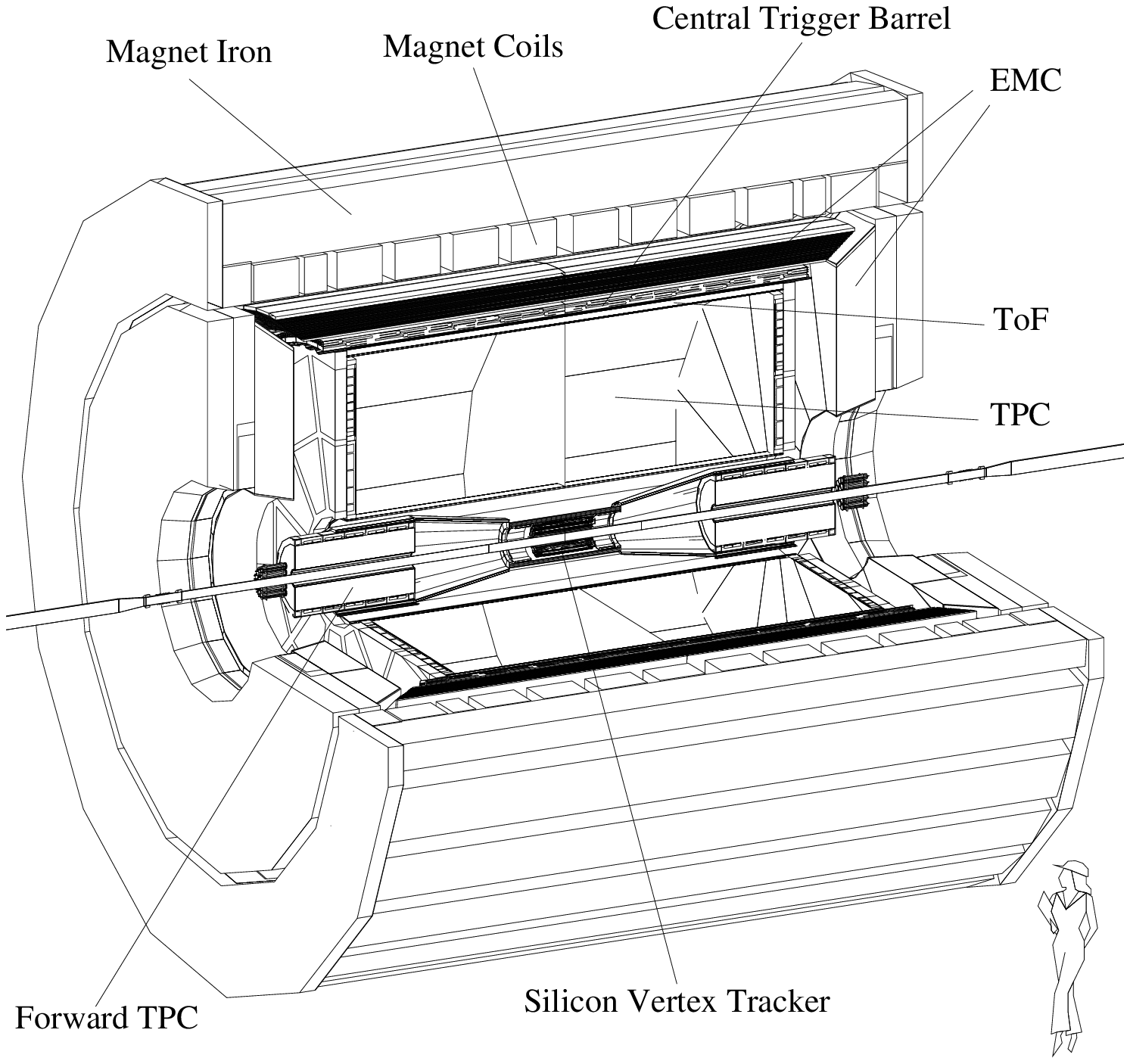}}
\captive{The STAR detector at RHIC. The outer radius of the TPC 
cylinder is 2~m and the length is 4~m.}
\label{STAR}
\end{figure}

The maximum 4--momentum transfer in a coherent nuclear interaction is 
determined by the nuclear form factor. The maximum momentum transfer for 
coherence is $Q_{max} \sim \hbar c / R$ in the rest 
frame of the nucleus, where R is the nuclear radius. This has two important 
consequences. First, the transverse momentum of the final state is limited to
$p_T \leq \sqrt{2} \hbar c / R$. Secondly, the maximum photon energy
is limited to $\gamma \hbar c / R$, where $\gamma$ is the Lorentz factor
of the nucleus in the relevant frame. 

The flux of virtual photons from the electromagnetic field of the nuclei
scales as $Z^2$, so the two-photon luminosity scales as Z$^4$. This is
one of the advantages of studying two--photon interactions in 
nuclear collisions. For Pomerons/mesons the scaling with A is more complicated 
because of nuclear shadowing, as will be discussed in the section on 
photonuclear interactions. 

The maximum two-photon center--of--mass energy is $\sim 2 \gamma \hbar c /R$, 
where $\gamma$ is the
Lorentz factor of the nuclei in the center--of--mass frame. This is
about 6~GeV for heavy nuclei at RHIC. RHIC will be the
first accelerator energetic enough to produce hadronic final states in 
coherent interactions.

\section{Two-Photon Interactions}

Two--photon interactions at RHIC will have energies in the region dominated
by resonance production. Neutral resonances
with spin J=0 or 2 can be produced in a collision of two quasi--real
photons. Meson pairs ($\pi^+ \pi^-$, $\pi^0 \pi^0$, $K^+ K^-$ etc.) and
lepton pairs ($e^+ e^-$, $\mu^+ \mu^-$, $\tau^+ \tau^-$) can also be 
produced. A few of the very many interesting physics topics that can be 
studied in two--photon collisions at RHIC are the following:

\noindent
{\bf Electron/Lepton pair production:}
Reactions of the type $Au+Au \rightarrow Au+Au+e^+e^-$. 
The calculation of the cross section for this process has 
attracted some interest recently\cite{Serbo98,Baltz98,Baur97,Baur95}. 
The process is pure QED, 
but $Z \alpha \sim$~0.6. Calculations
using first order perturbation theory violates unitarity (the interaction
probability becomes larger than 1) even at fairly large impact parameters
($b \sim \lambda_C = 386$~fm)\cite{Baur88}. 
Unitarity can be restored if higher order terms are included in the 
calculations\cite{Baur90}. These higher order terms lead to the production of 
multiple pairs. 
The importance of higher order effects is still under debate. Recent 
non--perturbative calculations have found that the $e^+e^-$ cross section
is identical to that obtained from lowest--order perturbation 
theory\cite{Baltz98}. Non--perturbative effects should be present
in the production of multiple pairs. These results were criticized in 
Ref.~\cite{Serbo98}, which concluded that higher order terms contribute
25\% of the $e^+e^-$ cross section at RHIC.
Measurements of single and multiple electron/positron pairs should
help resolving these theoretical difficulties and serve as a 
probe of strong--field QED. 

\noindent
{\bf Meson Spectroscopy:} 
Reactions of the type $\gamma \gamma \rightarrow$~Resonance can be used to
probe the quark content of the produced state. Photons couple to charge,
$\Gamma_{\gamma \gamma} \propto Q^4$. A nice illustration of this is the
two--photon widths of the neutral mesons (f$_2$(1270), a$_2$(1320), and 
f$_2$'(1525)) in the tensor meson nonet. 
These states are believed to have the quark compositions shown below. \newline
\begin {table} [h!] \begin{center}
\vspace{-0.7cm}
\begin{tabular} {lllll}
Meson & Quark Composition & $\Gamma{\gamma \gamma}$(relative) & 
$\Gamma{\gamma \gamma}$(measured)\\ 
 & & & & \\
f$_2$(1270) & $\frac{1}{\sqrt{2}} | u \overline{u} + d \overline{d} >$ &
25 & 2.8 keV\\
a$_2$(1320) & $\frac{1}{\sqrt{2}} | u \overline{u} - d \overline{d} >$ &
9 & 1.0 keV\\
f$_2$'(1525) & $| s \overline{s} >$ &
2 & 0.1 keV
\end{tabular}
\end{center}
\end{table}

\vspace{-0.8cm}
\noindent
From this one can easily compute the expected relative two--photon width. 
As can be seen, this is also in good agreement with the measured values.

For a pure glueball,
$\Gamma_{\gamma \gamma} = 0$ to first order and the cross section for 
two-photon production would vanish. 
Stringent limits on the two--photon widths is an important test for
glueball candidates.
Two of the most promising glueball candidates are the $f_0(1500)$\cite{ALEPH}
and the $f_J(2220)$ (also known as $\xi(2230)$)\cite{CLEO}.

\noindent
{\bf Meson Pair Production:} 
At the hadron level, photons couple only to charged mesons.  So, $\pi^+\pi^-$
should be produced, with $\pi^0\pi^0$ suppressed.  This picture 
applies near threshold, where the photon wavelength is large compared to the 
size of the meson.  However, at higher energies, the photons `see' and
couple to quarks and $\pi^+\pi^-$
and $\pi^0\pi^0$ are produced in comparable numbers.  By comparing the
rates of the two final states, the transition can be studied, and the
size of the mesons determined.  

However, for some channels, other mechanisms may apply.  The reaction
$\gamma\gamma\rightarrow\rho^0\rho^0$ is of special interest. A
resonance is observed near threshold for $M_{\rho\rho}\sim 1.3-1.6$
GeV.  A similar resonance is not observed in the $\rho^+\rho^-$
channel\cite{PDG}.

\begin{figure}
\label{lumfig}
\epsfxsize=0.85\textwidth
\centerline{\epsffile{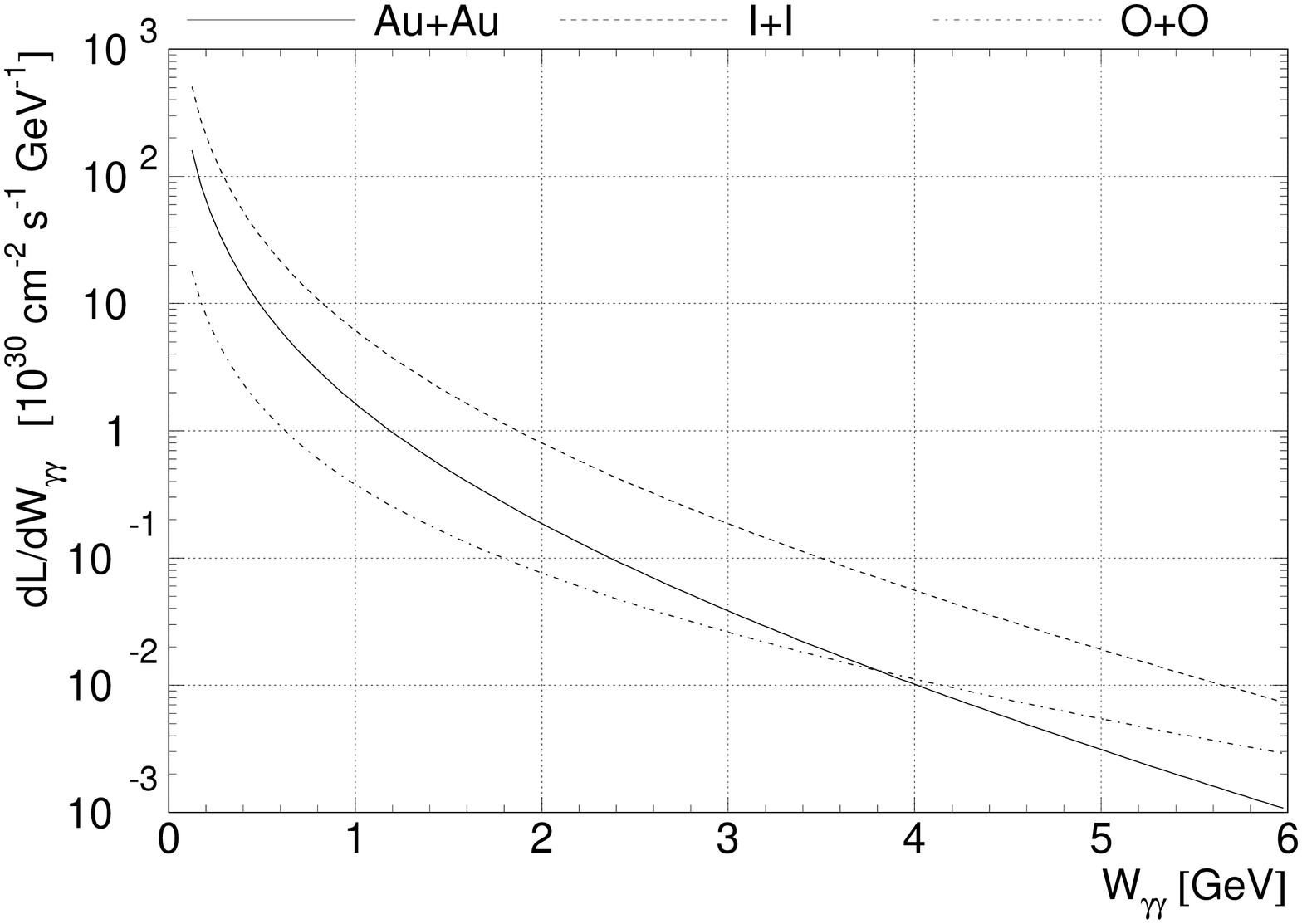}} 
\vspace{-0.5cm}
\captive{Differential two--photon luminosity, $d \cal{L}$/$dW$, for O+O,
I+I, and Au+Au collisions at RHIC.}
\end{figure}

\vspace{1mm}
The cross section for the two--photon production of a state 
is usually factorized into a cross section for 
$\gamma \gamma$ fusion and an equivalent $\gamma \gamma$ luminosity, 
\begin{equation}
   \sigma(A+A \rightarrow A+A+X) = \int \! \int 
   \frac{d {\cal L}_{\gamma \gamma}}{dW dY} 
   \sigma_{\gamma \gamma \rightarrow X}(W) dW dY \; .
\end{equation}
The two--photon cross sections, $\sigma_{\gamma \gamma \rightarrow X}(W)$, 
for single resonances, lepton and meson pairs can be found in 
Ref.~\cite{Brodsky}.

The two--photon luminosity in peripheral nucleus--nucleus collisions has 
been discussed by several authors \cite{AAgamma}. These 
analyses use the Weizs\-\"{a}cker--Williams method in the
impact parameter representation to calculate the equivalent photon flux.
When the nuclear impact parameter, $b$, is smaller than the sum of the nuclear
radii, hadronic interactions will dominate. To find 
the usable two-photon luminosity this region must be excluded 
from the integration. In the papers cited above, this is accomplished by 
introducing a sharp cut of at $b = R_1 + R_2$. The differential luminosity 
is 
\begin{equation}
   \label{lumeq}
   \frac{d{\cal L}_{\gamma \gamma}}{d \omega_1 d \omega_2} =
   {\cal L}_{A A} \int_{b_1>R} \int_{b_2>R} n(\omega_1,b_1)
   n(\omega_2,b_2) \Theta( \mid \vec{b}_1 - \vec{b}_2 \mid 
   - 2R) d^2b_1 d^2b_2 \; ,
\end{equation}
where $n(\omega,b)$ is the Weizs\-\"{a}cker--Williams photon flux from one
of the nuclei at a distance b from its center. The $\Theta$--function is
defined as $\Theta(x) = 0$ for $x<0$ and $\Theta(x) = 1$ for $x>0$.
Then, for $\gamma \gg 1$,
\begin{equation}
   n(\omega,b) =
   \frac{dN}{d \omega d^2b} = 
   \frac{Z^{2}\alpha}{\pi^2} \frac{1}{\omega b^2} x^2  K_1^2(x) \; .
\label{dndEdb}
\end{equation}
Here, $\omega$ is the photon energy, b the impact parameter, and 
$x = b \omega / \gamma$ \cite{Jackson}.
In Eq.~4 and in the rest of this paper, natural
units are used, i.e. $\hbar = c = 1$. Since $n(\omega,b)$ $\sim Z^2$, 
the luminosity essentially scales as $Z^4$. The scaling is inexact, however, 
because of the $\Theta$-function in Eq.~\ref{lumeq} and the variation of R 
with A/Z.

A variable transformation from the individual photon energies, $\omega_1$ and
$\omega_2$, to the $\gamma \gamma$ center--of--mass energy, $W$, and rapidity, 
$Y$, is achieved by
\begin{equation}
\label{kinematics}
\setlength{\arraycolsep}{0.2cm}
\begin{array}{rclcrcl}
\omega_1 &=& \frac{1}{2}We^Y    & \hspace*{1.5cm} & W &=& 
\sqrt{4 \omega_1 \omega_2}          \\[1.5mm]
\omega_2 &=& \frac{1}{2}We^{-Y} & \hspace*{1.5cm} & Y &=& 
\frac{1}{2}\ln(\omega_1/\omega_2) \; .\\
\end{array}
\end{equation}
The differential $\gamma \gamma$ luminosity, $d{\cal L}_{\gamma \gamma}/dW$, 
is then
\begin{equation}
   \frac{d{\cal L}_{\gamma \gamma}}{dW} = \int_{-\infty}^{\infty} 
   \frac{d{\cal L}_{\gamma \gamma}}{dW dY} dY \; . 
\end{equation}

\begin{figure}
\epsfxsize=0.7\textwidth
\centerline{\epsffile{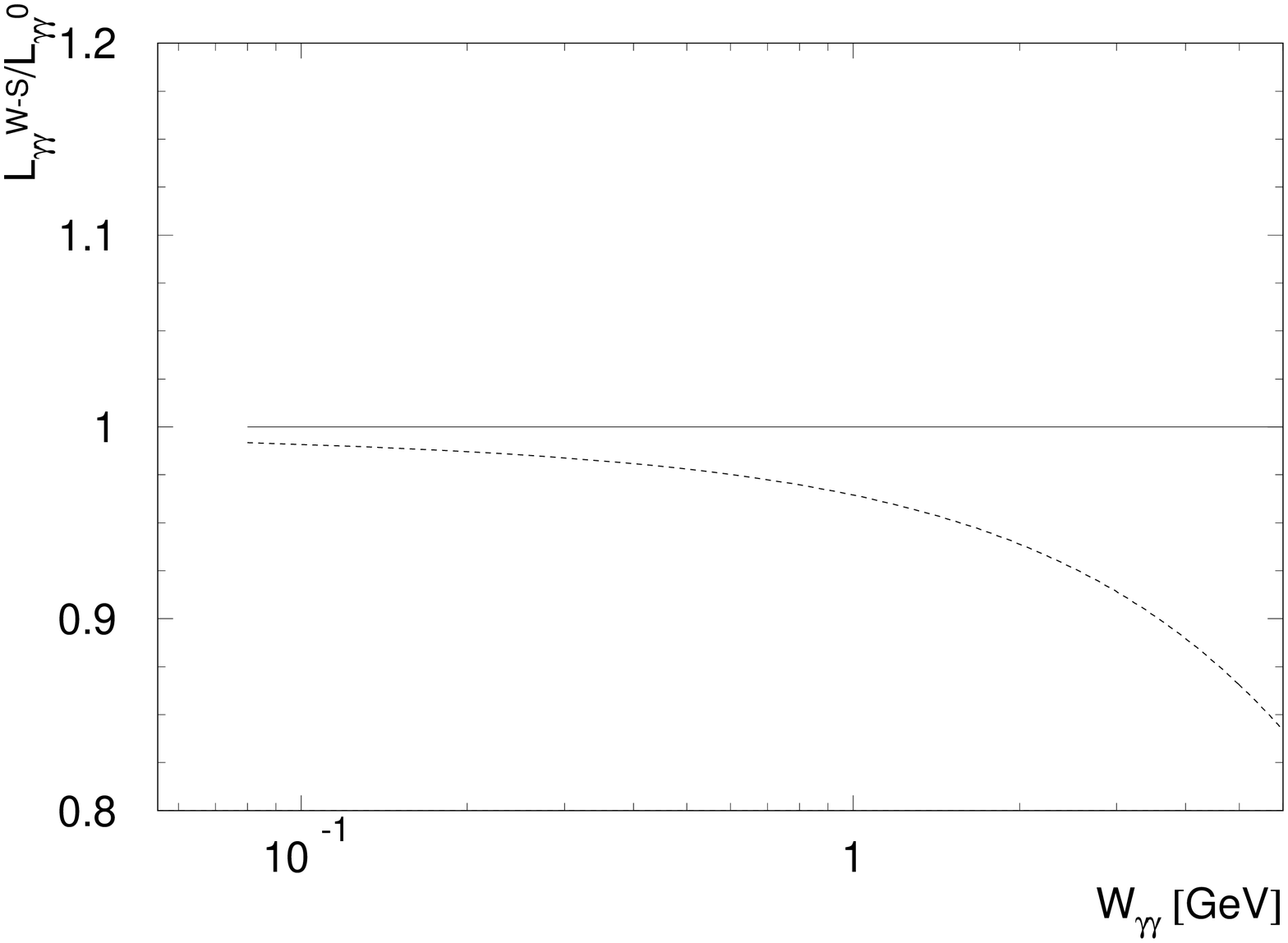}} 
\vspace{-0.5cm}
\captive{Reduction in two--photon luminosity when the nucleon density is
approximated with a Woods--Saxon distribution (dotted line) as compared 
with a flat distribution for $r<R_{nuc}$ (solid line). The calculation is
for gold--gold interactions at 100+100 A GeV.}
\label{wscomp}\end{figure}

Figure~2 shows the equivalent two--photon luminosity for three different 
nuclear systems at RHIC, calculated from Eq.~\ref{lumeq}.
The highest two--photon luminosity is obtained in I+I interactions. The 
lower $Z$ is compensated by higher nuclear luminosity and smaller nuclear 
radius. 

The sharp cut--off at $b = 2R$ in Eq.~\ref{lumeq} treats the nuclei as hard 
spheres. 
In a more realistic model one has to consider also the diffuseness of the 
nuclear surface. This can be accomplished by
first rewriting the expression for the luminosity as
\begin{equation}
   \label{wseq}
   \frac{d{\cal L}_{\gamma \gamma}}{d \omega_1 d \omega_2} =
   {\cal L}_{A A} \int_{b_1>R} \int_{b_2>R} n(\omega_1,b_1) \,
   n(\omega_2,b_2) \, [ \, 1 - P_{INT}( \mid \vec{b}_1 - \vec{b}_2 \mid) \, ] 
   \, d^2b_1 d^2b_2 \; . 
\end{equation}
where $P_{INT}(b)$ is the hadronic interaction probability for a 
nucleus--nucleus collision at impact parameter $b$. 
There is a finite probability for having an interaction also at impact
parameters $b>2R$. This probability can be calculated using
the Glauber Model:
\begin{equation}
\label{Glauber}
   P(b) = 1 - \exp \left( -\sigma_{nn} \int T_A(r) T_B(b-r) d^2r
   \right) 
\end{equation}
where $\sigma_{nn}$ is the hadronic nucleon--nucleon cross section and $T(r)$ 
is the nuclear thickness function\cite{Glauber}. Here, $\sigma_{nn}=$~52~mb 
is used, corresponding to $\sqrt{s}$=200~GeV \cite{PDG}. The 
thickness function is calculated from the nucleon density distribution $\rho$,
\begin{equation}
\label{T}
   T(\vec{b}) = \int \rho(\vec{b},z) dz \; .
\end{equation}
For $\rho$, a Woods--Saxon or Fermi distribution is used:
\begin{equation}
   \rho(r) = \frac{\rho_0}{1 + \exp( \frac{r - R_{nuc}}{c} )} \; .
\end{equation}
where $R_{nuc}$ is the nuclear radius and c is the skin--thickness. Here,
the radius is calculated from $R_{nuc} = r_0 A^{1/3}$ with $r_0 = 
1.16(1.-1.16A^{-2/3})$~fm, and a constant skin--thickness of c=0.53~fm is 
used. 
These parameterizations are obtained from electron--nucleus scattering 
data\cite{nucsize}.

The effect of using a Woods--Saxon distribution
for the nucleon density as compared with a $\Theta$--function is
shown by the dotted curve in Fig.~3. The luminosity
is reduced up to 15\%, depending on energy.

The cross sections and production rates (using a Woods--Saxon distribution) for
various resonances, muon and tau pairs, and $\rho \rho$ pairs are shown in 
Table~2 for Au+Au and I+I interactions. For mesons in the mass 
range 0--2~GeV, the production rates will be between $10^5$--$10^7$ events
per year. In section 5 these rates will be compared with the rates of 
background reactions. 

\begin{center}
\begin {table} [t!] \begin{center}
\begin{tabular} {lrrcrc} \hline
\hspace*{0.0cm} Final State & $\Gamma_{\gamma \gamma}$  & 
\multicolumn{2}{c}{Au+Au}  & \multicolumn{2}{c}{I+I}    \\ 
         &   $[$keV$]$      & $\sigma$ $[\mu$b$]$ & Evts./Year & 
$\sigma$ $[\mu$b$]$ & Evts./Year \\ \hline  
$\pi^0$  &  7.8 eV   &  4700 & 9.4 $\cdot$ 10$^{6}$  & 1100 & 
3.0 $\cdot$ 10$^{7}$  \\
$\eta$   &  0.5      &   884 & 1.8 $\cdot$ 10$^{6}$  &  229 & 
6.2 $\cdot$ 10$^{6}$ \\ 
$\eta '$ &  4.3      &   642 & 1.3 $\cdot$ 10$^{6}$  &  178 & 
4.8  $\cdot$ 10$^{6}$ \\
$f_0(980)$ & 0.6  &    75 & 1.5  $\cdot$ 10$^{5}$  &   21 & 
5.6 $\cdot$ 10$^{5}$ \\
$f_2(1270)$ & 2.8 &   514 & 1.0 $\cdot$ 10$^{6}$  &  149 & 
4.0 $\cdot$ 10$^{6}$ \\ 
$a_2(1320)$ & 1.0 &   155 & 3.1 $\cdot$ 10$^{5}$  &   45 & 
1.2 $\cdot$ 10$^{6}$ \\
$f_2'(1525)$ & 0.1  &     7 & 1.4 $\cdot$ 10$^{4}$  &    2 & 
5.8 $\cdot$ 10$^{4}$ \\
$\eta_c(2980)$ & 7.5 &    2 & 4.7 $\cdot$ 10$^{3}$  &    1 & 
2.3 $\cdot$ 10$^{4}$ \\ \hline
$\mu^+ \mu^-$ & -- & 130 mb & 2.7 $\cdot$ 10$^{8}$  &   33 mb &
8.9 $\cdot$ 10$^{8}$ \\
$\tau^+ \tau^-$ & -- & 0.65 & 1.3 $\cdot$ 10$^{3}$  & 0.29  &
8.0 $\cdot$ 10$^{3}$ \\ \hline
$\rho^0 \rho^0$ & -- & 20   & 4.0 $\cdot$ 10$^{4}$ & 6  &
1.6 $\cdot$ 10$^{5}$ \\ \hline
\end{tabular}
\label{prodrates}
\captive{Cross sections and production rates at RHIC for various final states
in two--photon interactions at design luminosity. One year corresponds to 
$10^7$ seconds of operation. The $\rho^0 \rho^0$ is in the invariant
mass range 1.5--1.6~GeV.}
\end{center}
\end{table}
\end{center}

\section{Photonuclear Interactions}

In this section exclusive vector meson production in photonuclear interactions
will be discussed. These reactions are of the type 
$\gamma + A \rightarrow V + A$, where $V$ is a vector meson ($\rho$, 
$\omega$, $\phi$ and J/$\Psi$ are discussed here). The reactions
are usually described within the framework of the Vector Dominance Model and
are assumed to proceed in the following way: One of the nuclei
emits a photon, which fluctuates into a vector meson. The vector meson
then scatters off the other nucleus. 

\begin{figure}[t!]
\epsfxsize=0.5\textwidth
\centerline{\epsffile{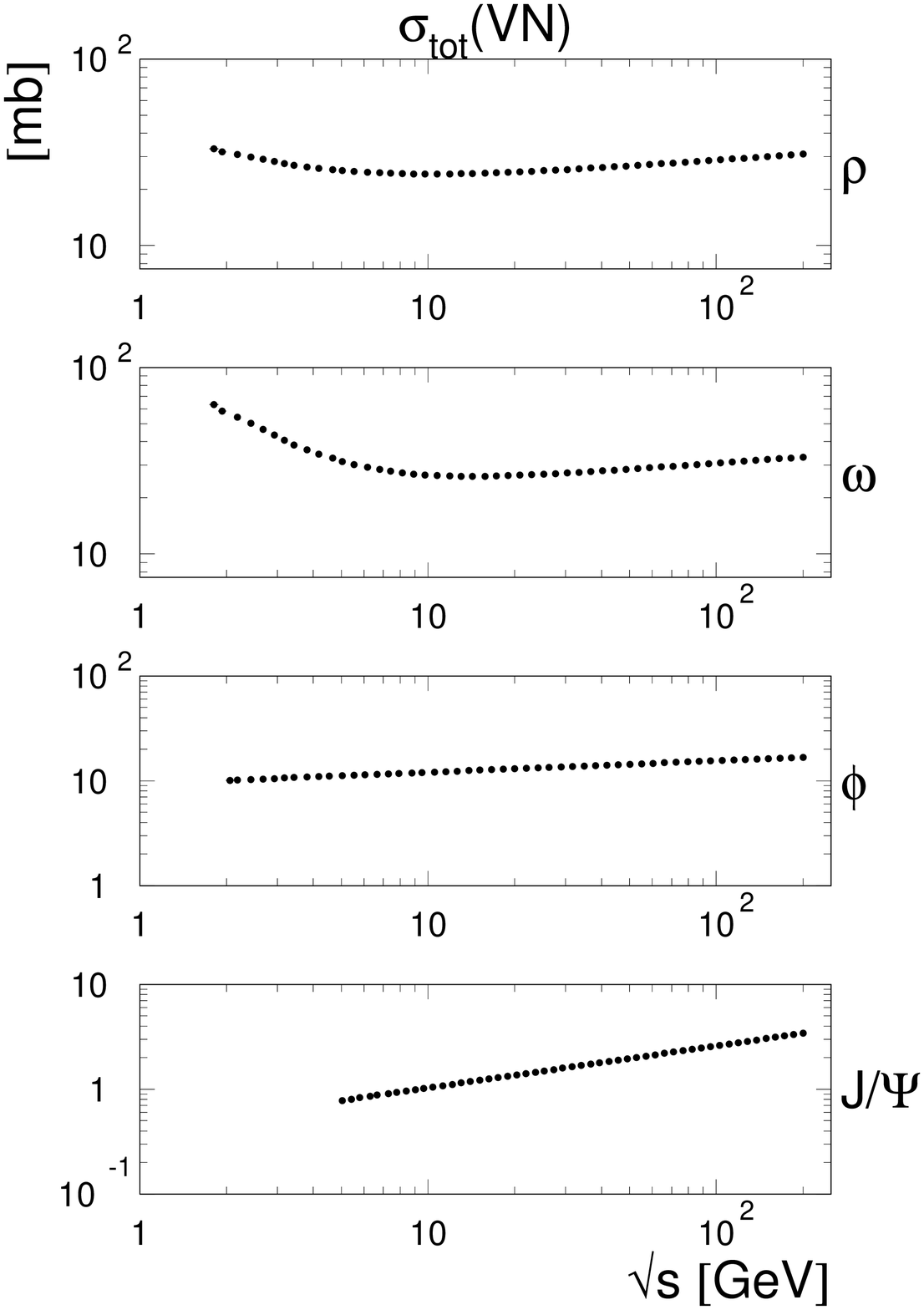}} 
\vspace{-0.5cm}
\captive{Total inelastic vector meson--nucleon cross sections.}
\label{sigvn}
\end{figure}

There are several reasons to study these interactions in nuclear systems.
The reaction $\gamma + N \rightarrow V + N$ (N is a nucleon) is mediated by
mesons (dominant at low energy) and by Pomerons (dominant at high energy).
By extending these studies to nuclei one can study how Pomerons and mesons
couple to nuclei. 

Photonuclear J/$\Psi$ production is of special interest
since in  photonucleon interactions the energy dependence violates the 
soft--Pomeron/Regge-theory behaviour. It has been suggested that the 
diffractive production of J/$\Psi$--mesons on nuclear targets can probe the
gluon distributions in nuclei\cite{Brodsky94,Ryskin97}.

The cross section for the exclusive production of one vector meson
is calculated as the product of the Weizs\"{a}cker--Williams photon
flux, $n(\omega)$, with the photonuclear cross section, $\sigma_{\gamma A}$,
integrated over photon energy,
\begin{equation}
   \sigma(A+A \rightarrow A+A+V) = 2 \int \sigma_{\gamma A \rightarrow V A}
   (\omega) \, n(\omega) \, d \omega \;.
\end{equation}
The factor of 2 is because either nucleus can emit a photon.
The photon flux is calculated by integrating Eq.~\ref{dndEdb} over all 
impact parameters $b>2R$ , and averaging over the area of the target 
nucleus\cite{Baur93}.

The 4--momentum transfer from the target nucleus, $t$, will be determined
by the nuclear form factor, $F(t)$, in coherent reactions. The cross section is
\begin{equation}
  \sigma_{\gamma A}(E_{\gamma}) = \int_{t_{min}}^{\infty} \left. 
  \frac{d\sigma}{dt} \right|_{t=0} \left| F(t) \right|^2 dt \; .
\end{equation}
$F(t)$ can be approximated with a gaussian,
\begin{equation}
     F(t) =  e^{-|t|/2Q_{0}^{2}} \; .
\end{equation}
\begin{center}
\begin {table} [b!] \begin{center}
\begin{tabular} {crrrrrc} \hline
Meson    & $f_v^2/4 \pi$ & $X$ $[\mu$b$]$ & \multicolumn{1}{c}{$\epsilon$} & 
Y $[ \mu$b$]$ & \multicolumn{1}{c}{$\eta$} & $b$ $[$GeV$^{-2}]$ \\ \hline 
$\rho^0$ &  2.02         & 5.0 &    0.22    & 26.0  & 1.23     & 11  \\
$\omega$ & 23.13         & 0.55&    0.22    & 18.0  & 1.92     & 10  \\
$\phi$   & 13.71         & 0.34&    0.22    & --    & --       &  7  \\
$J/\Psi$ & 10.45         & 0.0015&  0.80    & --    & --       &  4  \\ \hline
\end{tabular}
\label{Y}
\captive{Photon--vector meson couplings and parameterizations
of the vector meson cross sections. The vaules for $b$ 
are from Ref.~\cite{Crittenden}. See text for details.} 
\end{center}
\end{table}
\end{center}
For a gold nucleus $Q_0=60$~MeV \cite{Ellis}. In the narrow width 
approximation ($\Gamma_V = 0$),  
$t_{min} = [M_V^2/4 \omega \gamma]^2$. A more realistic model includes the 
natural width of the vector meson (important for the $\rho$)\cite{Prep}.

\begin{figure}
\epsfxsize=0.85\textwidth
\centerline{\epsffile{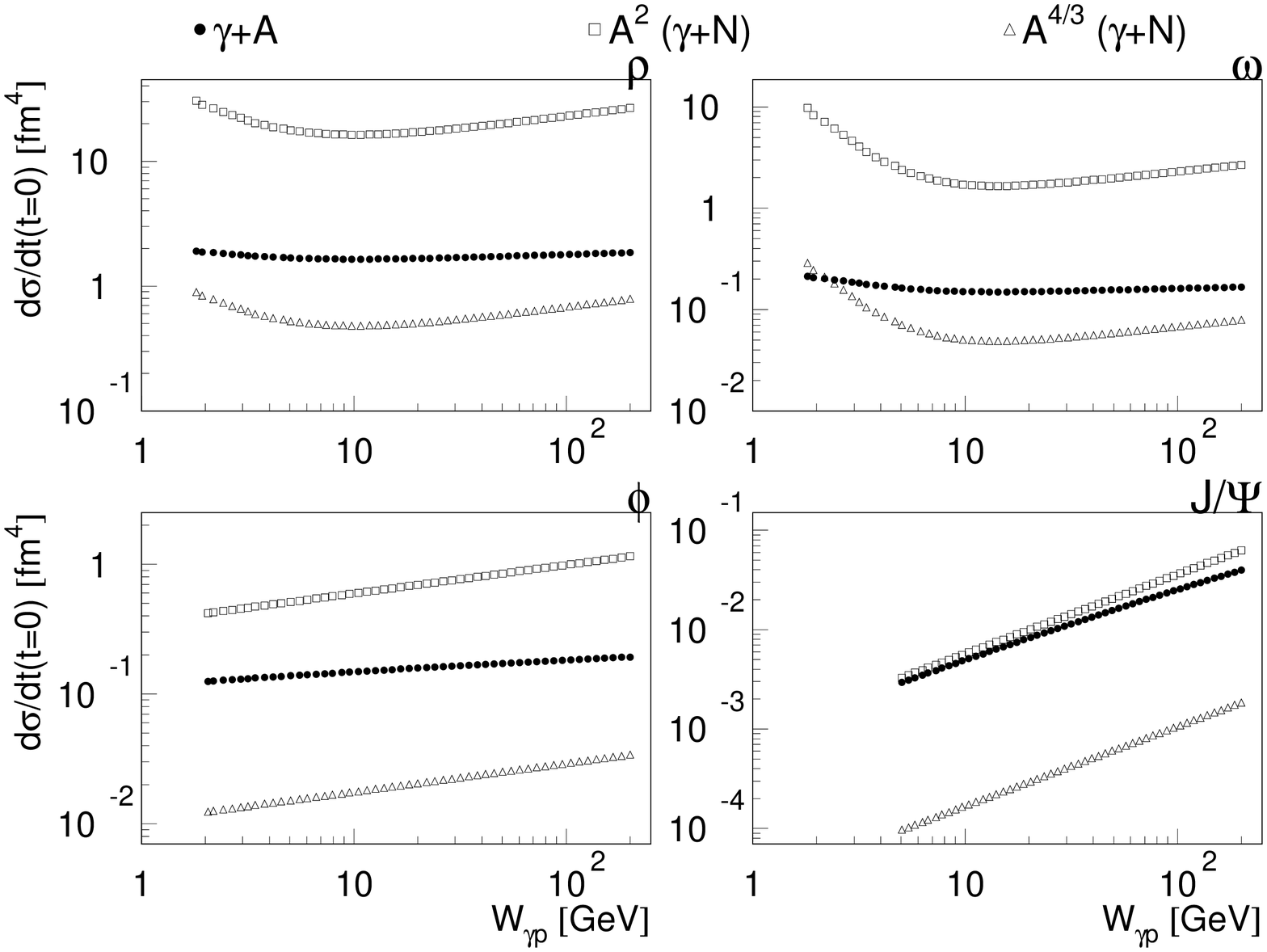}} 
\vspace{-0.5cm}
\captive{Forward scattering amplitudes for diffractive vector meson 
production on gold nuclei.}
\label{dsigdt}
\end{figure}

The forward scattering amplitude is extrapolated from $\gamma$N
interactions. Studies at the 
electron--proton collider HERA 
have shown that the cross section for exclusive vector meson production
at high energies increases slowly with energy. The \pagebreak variation is 
consistent 
with Regge theory and soft--Pomeron exchange. 
Exclusive vector meson production is summarized in Ref.~\cite{Crittenden}, 
from which the following parameterization can be extracted:
\begin{equation}
\label{sigma}
   \sigma(\gamma p \rightarrow Vp) \, = \, X \, 
{\scriptstyle W}_{\gamma p}^{\, \epsilon} \, 
   + \, Y \, {\scriptstyle W}_{\gamma p}^{- \eta} \; , 
\end{equation}
where ${\scriptstyle W}_{\gamma p}$ is the $\gamma p$ center--of--mass energy
(in GeV).
The values of $X$, $Y$, $\epsilon$, and $\eta$ are given in Table~3.
For $\phi$ and J/$\Psi$ the cross section rises monotonically
with $s$ and only the first term in Eq.~\ref{sigma} is necessary. 
For these mesons only Pomeron exchange contribute\cite{Freund}, while for 
the $\rho$ and $\omega$ meson exchange dominates at low energies. 

The $t$ dependence can be parameterized as
\cite{Crittenden,Zeus98}
\begin{equation}
   \frac{d \sigma}{dt} = \left. \frac{d \sigma}{dt} \right|_{t=0} 
   e^{-b|t|+c|t|^2} \; .
\end{equation}
The constant $c$ is small, and the behaviour is essentially exponential. To
simplify the following calculations, it will be assumed that $c=0$. The 
forward scattering amplitude is then
\begin{equation}
   \left. \frac{d \sigma}{dt} \right|_{t=0} = 
        b \cdot ( X \, {\scriptstyle W}_{\gamma p}^{\, \epsilon}  +  
        Y \, {\scriptstyle W}_{\gamma p}^{- \eta} ) \;.
\end{equation}

\begin{figure}
\epsfxsize=0.85\textwidth
\centerline{\epsffile{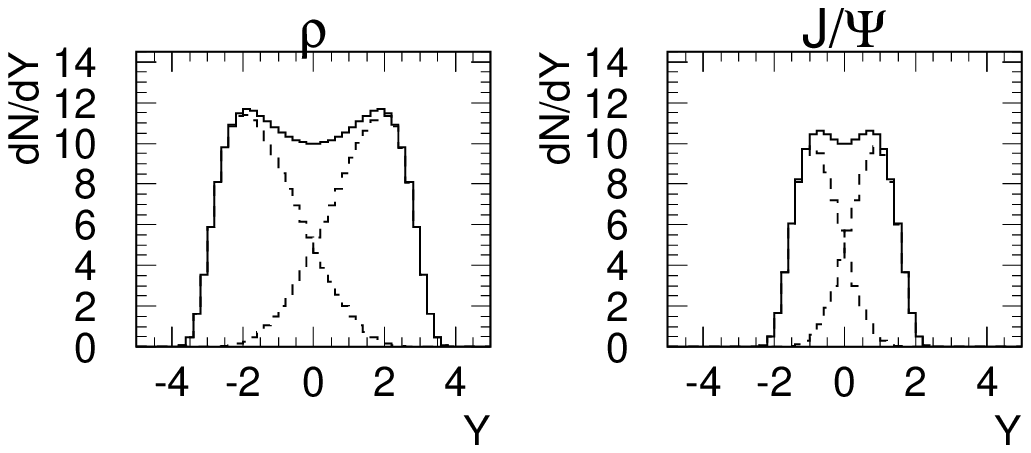}} 
\vspace{-1.5cm}
\captive{Rapidity distributions of $\rho$ and J/$\Psi$ mesons produced in 
exclusive reactions $Au+Au \rightarrow Au+Au+V$. The total distribution (solid
histogram) is the sum of the distributions for photon scattering off each of 
the projectile nuclei (dashed histograms).}
\label{evmdndy}
\end{figure}

In Vector Meson Dominance, the forward scattering amplitude for the reaction 
$\gamma + N \rightarrow V + N$ is given by the forward scattering 
amplitude for elastic $V+N$ scattering times the $\gamma$--vector meson 
coupling. The coupling is related to the width for the decay into 
an $e^+ e^-$ pair, 
\begin{equation}
\label{coupling}
   \frac{f_v^2}{4 \pi} = \frac{1}{3} 
                         \frac{m_v \alpha^2}{\Gamma_{V \rightarrow e^+ e^-}} 
   \; ,
\end{equation}
where $m_v$ is the vector meson mass\cite{Donnachie}. Using the values of 
$\Gamma_{V \rightarrow e^+ e^-}$ in Ref.~\cite{PDG}, the couplings in 
Table~3 are obtained. Eq.~\ref{coupling} is modified slightly when 
higher vector meson states are taken into account (Generalized Vector Meson 
Dominance Model), but this is not considered here. 
The total inelastic vector meson--nucleon cross section can then be related 
to the forward scattering amplitude by the optical theorem,
\begin{equation}
   \frac{\sigma_{tot}^2}{16 \pi} = \frac{f_v^2}{4 \pi \alpha} 
   \left. \frac{d\sigma}{dt} \right|_{t=0} (\gamma N \rightarrow V N) \; .
\end{equation}
The inelastic cross sections obtained from this are shown in 
Fig.~4. The 
cross sections for $\rho$ and $\omega$ are of the same order of magnitude as 
those for
$\pi$--nucleon scattering, whereas the cross section for $\phi$-- and 
J/$\Psi$--nucleon scattering are smaller. 

The total inelastic cross sections determine the scaling of $d \sigma / d t$ 
when going from $\gamma$--nucleon to $\gamma$--nucleus scattering 
through the optical theorem:
\begin{equation}
   \frac{\left. d \sigma / dt \right|_{t=0} (\gamma A \rightarrow V A)}
        {\left. \, \, d \sigma / dt \right|_{t=0} (\gamma N \rightarrow V N)} =
   \frac{\sigma_{tot}^2(VA)}{\sigma_{tot}^2(VN)}
\end{equation}
The total inelastic vector meson--nucleus cross sections are calculated
from the inelastic cross sections for vector meson--nucleon scattering using
the Glauber Model (cf. Eqs.~\ref{Glauber} and \ref{T}):
\begin{equation}
   \sigma_{tot}(VA) = \int (1 - e^{- \sigma(VN) T(b)}) d^2b \; .
\end{equation}

The resulting forward scattering amplitudes are plotted in Fig.~5.
The calculated values for $\gamma$-nucleus scattering are compared with a
simple $A^{4/3}$ and $A^2$ scaling of the corresponding value for 
$\gamma$--nucleon scattering. 
The scaling $A^2$ corresponds to weak absorption ($\sigma_{VN}$ small),
and $A^{4/3}$ corresponds to strong ``black disc'' absorption 
($\sigma_{VN}$ large).
For the $\rho$ and the $\omega$, the scaling is closest
to $A^{4/3}$. For the J/$\Psi$ nuclear shadowing is less important because
of the much smaller inelastic cross section, and the scaling is close to $A^2$.
The $\phi$ shows more shadowing than the J/$\Psi$ but less than the
lighter mesons. The unit for $d \sigma / d t$ is $[ \, \mu$b GeV$^{-2}]$. 
For simplicity, this has been converted to $[fm^4]$ in the figure.

The cross sections for gold--gold interactions are listed in 
Table~4. The table also lists the corresponding production 
rates at the design luminosity. 
Both the cross sections and production rates are high. The cross section for
diffractive $\rho$--meson production is roughly 10\% of the total inelastic
Au+Au cross section. The rates are higher than for most two--photon final 
states. 

The rapidity distributions for $\rho$ and J/$\Psi$ final states are shown
in Fig.~6. The narrower distribution for the J/$\Psi$ is a
kinematical effect caused by the higher mass of the J/$\Psi$.

\begin{center}
\begin {table} [h!] \begin{center}
\begin{tabular} {lrrr} \hline
Meson    & $\sigma$ $[$mb$]$ & Prod. Rate $[$Hz$]$ \\
\hline 
$\rho^0$ &   640             & 130    \\
$\omega$ &    60             &  12    \\
$\phi$   &    41             &   8    \\
$J/\Psi$ &   320 $\mu$b      &  6.5 $\cdot 10^{-2}$ \\ \hline
\end{tabular}
\label{vmrates}
\captive{Cross sections and production rates for exclusive vector meson 
production in Au+Au interactions.}
\end{center}
\end{table}
\end{center}

\vspace{-0.6cm}
\section{Experimental Techniques}

Two--photon and coherent photonuclear interactions have never been studied
in heavy--ion collisions before. 
To be able to
study them it is necessary to develop adequate 
triggering and analysis techniques. This section will discuss the 
experimental procedures that have been developed for the STAR experiment. 

It will not be possible to tag the outgoing nuclei, because
the typical transverse momentum transfers ($\approx 30$~MeV for Au) 
are completely negligible compared with the total
energy of the nuclei (19.7~TeV for Au); the resulting angular deflections will 
be of the order of microradians. 
Other techniques must be used to separate the coherent interactions 
from background processes. 

The analysis is based on using the characteristics of coherent events and 
developing cuts that reject as much of the background as possible. It is
assumed that the entire final state is detected and that no other particles 
are present in the event. The following variables are used:

\begin{Enumerate}

\item \underline{Charged Particle Multiplicity:} 
The resonances formed in two--photon interactions decay into final states
with low multiplicity (2 or 4 charged particles for the states studied here).

\item \underline{Charge Conservation:} Since no charge exchange occur the 
final state is required to obey $\sum Q_i = 0$.

\item \underline{FTPC Multiplicity:} It is required that the entire 
event is detected in the main TPC. By rejecting events with tracks in the 
forward TPCs backgrounds are reduced.

\item \underline{Final State Rapidity:} Two--photon events are centered 
around midrapidity
with a fairly narrow width. Requiring $| y_{cm} | <$~0.75 reduces the 
background, particularly from beam--gas and incoherent photonuclear 
interactions, with little loss of signal.

\item \underline{Transverse Momentum:} Coherent events 
will have $p_T \sim 1/R$. Cuts on the transverse momenta of 
$p_T \leq$~100~MeV/c and $p_T \leq$~50MeV
are used for triggering and off--line analysis, respectively.

\end{Enumerate}

Two--photon events have been generated with the STARLight Monte 
Carlo\cite{STARLight}. The geometrical acceptance of the main TPC has been 
conservatively defined as $| \eta | <$~1.5 and $p_T >$~150~MeV with 100\% 
detection efficiency. For the forward TPC the corresponding cuts are 
$2.5 \leq \mid \eta \mid \leq 3.75$ and $p_T >100$~MeV/c.

The following sources of backgrounds have been identified:
peripheral (hadronic) nucleus--nucleus collisions,
beam--gas interactions, upstream interactions, photonuclear 
interactions (incoherent), and 
cosmic rays. The backgrounds have been simulated with different Monte Carlo
codes. For hadronic nucleus--nucleus collisions and beam--gas events, 
FRITIOF~7.02\cite{FRITIOF} and VENUS~4.12\cite{VENUS} have
been used. Photonuclear events have been simulated with 
DTUNUC~2.0\cite{DTUNUC}.
For cosmic rays (muons) HemiCosm\cite{HemiCosm} has been used.
Details of these simulations are given in 
Ref.~\cite{starnote}. Here, the general techniques will be described together
with the current best background estimates. The simulations are
focussed on two--photon final states. 
The backgrounds to photonuclear events have not been studied in detail yet.
However, because of the similar kinematics, the backgrounds are expected to 
be similar to $\gamma \gamma$ reactions, while the signals are considerably
higher. Simulations have been performed both for the trigger and the off--line
analysis. The trigger will be considered first. 

The STAR trigger has 4 levels (0, 1, 2, and 3).
Levels 1 and 2 are combined here because it is not clear what
calculations are possible in Level 1 and what will have to wait until
Level 2. Level 0 uses programmable logic to determine the multiplicity in
$\sim 2 \mu s$. Inputs are the CTB and the TPC wire multiplicity, 
$| \eta | <$~2. 
Levels 1 and 2 use the same information, processed by a computer for more
accurate multiplicity information (e.g. merging adjacent hits).  
TPC tracking will be available in Level 3, after approximately 10~ms. 

In Level~0 a multiplicity cut of $2 \leq n_{ch} \leq 5$ is applied in 
combination with a crude topology cut (back--to--back in $\phi$). For 
Levels 1/2 the multiplicity cut is sharpened to $n_{ch} = 2$ or 4, and
it is required that there are no tracks in the region $1.5 < | \eta | < 2.0$.
Level 3 requires $p_T \leq 100$~MeV/c and the vertex is 
required to be within the reaction diamond. 
The background trigger rates after these cuts have been applied are shown 
in Table~5. Even at Level 1/2 these background rates are less than the signal
rates from photonuclear interactions. The rates are well within the 
capabilities of the STAR trigger system.

\vspace{-0.5cm}
\begin{center}
\begin {table} [h!] \begin{center}
\small
\begin{tabular} {cccccccc} \hline
Trigger         & \multicolumn{2}{c}{Hadronic A+A} & \multicolumn{2}{c}{Beam--Gas} &
$\gamma$+A       & Cosmic Rays & Total\\
Level    & FRITIOF & VENUS & FRITIOF & VENUS & DTUNUC & HemiCosm & Rate \\ \hline
0        & 18      & 21    & 53      & 53    &  63    &  26      & 160 \\
1/2      &  2      &  4    &  6      &  7    &  11    &  18      &  38\\
3        &  0.03   &  0.1  &  0.1    &  0.1  &   0.6  &   0.4    & 1.2 \\ \hline
\end{tabular}
\label{trigger2}
\captive{Trigger Rates (in Hz) of the various background processes discussed 
in the text for gold--beams at the design luminosity.}
\end{center}
\end{table}
\end{center}

Detailed analyses have been performed for the following three systems:
one intermediate mass meson, $f_2(1270) \rightarrow \pi^+ \pi^-$;
one meson pair, $\rho \rho \rightarrow \pi^+ \pi^- \pi^+ \pi^-$; and
one heavy meson, $\eta_c(2980) \rightarrow K^+ \pi^- K^- \pi^+$.
These systems are representative of a wide selection of two--photon
final states. 

The production rates for these states are shown in Table~2. 
Taking into account the branching ratios for the decay modes studied and the
geometrical acceptance of STAR gives the rates shown on the first line in
Table~6. Applying the cuts 1--5 for the peripheral collisions analysis
above gives the rates shown on the second line of the same table. 
A similar analysis was made on the background 
events generated with FRITIOF/VENUS/DTUNUC, and the results are shown in the 
table. For the backgrounds, suitable cuts have been applied 
on the invariant mass. 
As an illustration, the effect on the transverse momentum distribution
of the cuts 1--4 above is shown in Figure~7. 

\begin{figure}[t!]
\epsfxsize=0.65\textwidth
\centerline{\epsffile{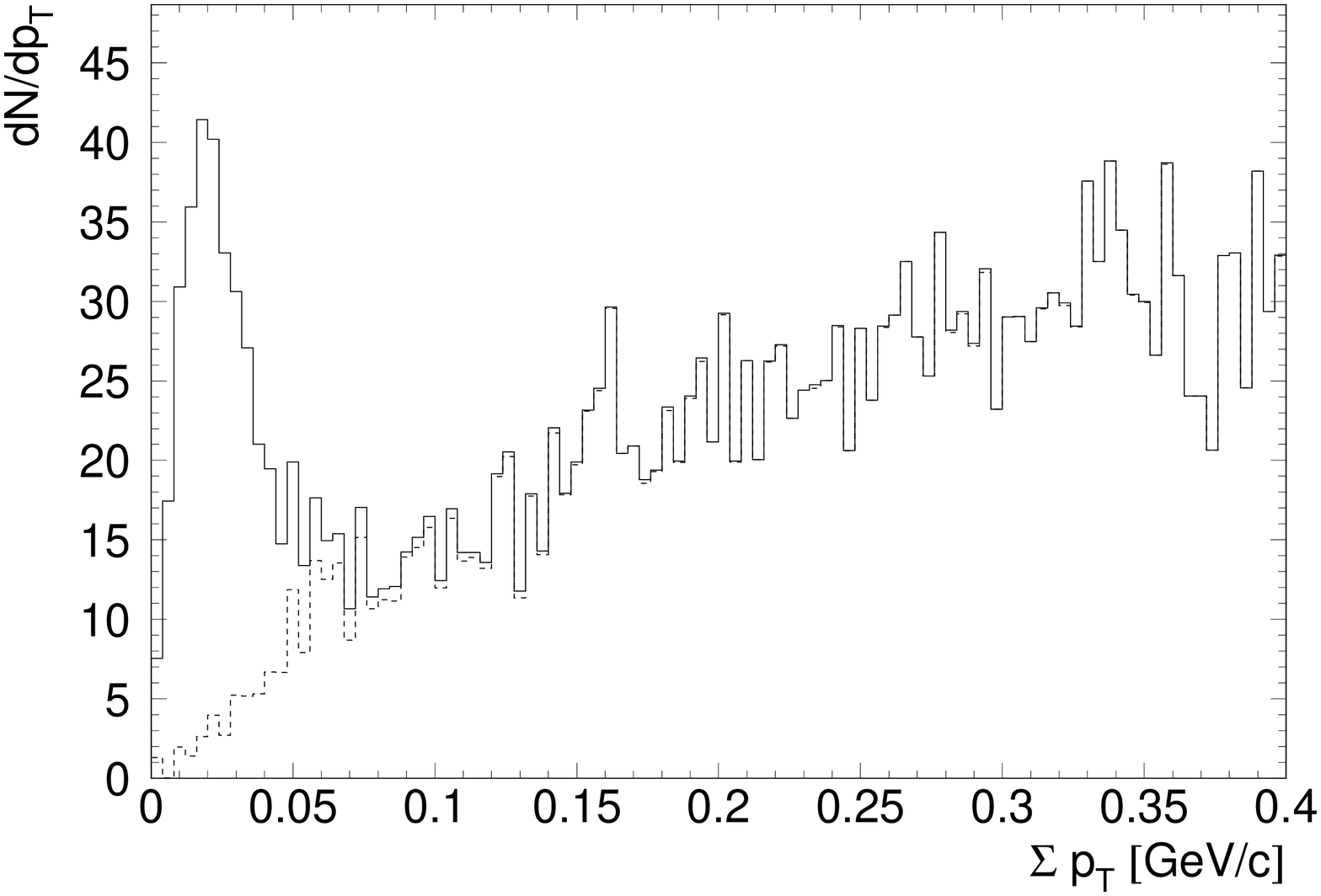}} 
\vspace{-0.5cm}
\captive{Transverse momentum distribution for $\gamma \gamma \rightarrow 
f_2(1270) \rightarrow \pi^+ \pi^-$ and backgrounds. The solid histogram 
shows the signal plus backgrounds and the dashed histogram shows the 
backgrounds.}
\end{figure}

As can be seen, the signal to noise ratio will be very high for $f_2(1270)$ and
$\rho$--pairs. For the $\eta_c$ particle identification (separation of pions
from kaons) is necessary in order to suppress the backgrounds sufficiently. 

\begin{center}
\begin {table} [h!] \begin{center}
\begin{tabular} {lrrr} \hline
                             & \multicolumn{3}{c}{System}           \\ 
                             & $f_2(1270)$ & $\rho^0\rho^0$ & $\eta_c(2980)$ \\
\hline
Within STAR Acceptance       & 380,000        &  9,500       & 33 \\
After analysis cuts          & 300,000        &  8,500       & 30 \\ \hline
Peripheral AA (FRITIOF/VENUS)&     400/4,000  &    50/460    & 90/680  \\
Beam--gas (FRITIOF/VENUS)    &     400/1,000  &     6/6      & $<$1/$<$1  \\
$\gamma$+A (DTUNUC)          &  15,000        &  1,000       & 730 \\ \hline
Background without PID    &  16,000/20,000 &  1,000/1,500 & 820/1,400 \\ \hline
Background with PID       &   	       &  	      & 9/15 \\ \hline
\end{tabular}
\label{fs}
\captive{Estimated signal and background rates (Events/year) with and
without particle identification, for two--photon production of $f_2(1270)$,
$\rho^0\rho^0$ and $\eta_c(2980)$ in Au+Au collisions.}
\end{center}
\end{table}
\end{center}

\section{Conclusions}

RHIC will be the first heavy--ion accelerator energetic enough to produce
final states with masses of several GeV in coherent interactions.
This will provide an 
opportunity for the study of many interesting physics topics. Some of these,
e.g. strong--field multiple $e^+ e^-$--pair and coherent vector meson 
production, are unique to heavy--ion colliders.

The rates for coherent two--photon and photonuclear interactions have been
calculated and are found to be high. The experimental feasibility of studying
these interactions have been demonstrated. Analysis techniques and 
algorithms have been developed which allow the separation of signal from
background.

\vspace{0.5cm}
\noindent
{\Large\bf Acknowledgements}

\vspace{0.3cm}
\noindent
J.N. would like to thank the organizers for the invitation to give a talk 
at this enjoyable conference at his {\it Alma Mater}.
This work was supported by the U.S. DOE under contract DE--AC--03--76SF00098.


\vspace{-0.3cm}

\begin{thebibliography}{99}

\bibitem{RHIC}
Conceptual Design of the Relativistic Heavy Ion Collider,
BNL--52195 (Brookhaven National Laboratory, May 1989).

\bibitem{STAR}
STAR Collaboration, Conceptual Design Report for the Solenoidal Tracker 
at RHIC,
LBL PUB--5347, June 1992.

\bibitem{Pomeron}
Quantum Chromodynamics and the Pomeron, J.R.~Forshaw and D.A.~Ross 
(Cambridge University Press, 1997).

\bibitem{Schramm}
A.J.~Schramm and D.H.~Reeves,
Phys. Rev. {\bf D 55} (1997) 7312.

\bibitem{Serbo98}
D.Yu.~Ivanov, A.~Schiller, and V.G.~Serbo,
hep--ph/9809281.

\bibitem{Baltz98}
A.J.~Baltz and L.~McLerran,
Phys. Rev. {\bf C 58} (1998) 1679;
B.~Segev and J.C.~Wells,
Phys. Rev. {\bf A 57} (1998) 1849;
U.~Eichmann, J.~Reinhardt, and W.~Greiner, \mbox{nucl--th}/9806031;
U.~Eichmann, J.~Reinhardt, S.~Schramm, and W.~Greiner, 
\mbox{nucl--th}/9804064 (to appear in Phys. Rev. {\bf A}).

\bibitem{Baur97}
A.~Alscher, K.~Hencken, D.~Trautmann, and G.~Baur,
Phys. Rev. {\bf A 55} (1997) 396.

\bibitem{Baur95}
K.~Hencken, D.~Trautmann, and G.~Baur,
Phys. Rev. {\bf A 51} (1995) 998; {\it ibid.} {\bf A 51} (1995) 1874.

\bibitem{Baur88}
C.A.~Bertulani and G.~Baur,
Phys. Rep. {\bf 163} (1988) 299.

\bibitem{Baur90}
G.~Baur, Phys. Rev. {\bf A 42} (1990) 5736.

\bibitem{ALEPH}
ALEPH Collaboration, Proc. EPS--HEP Conference, Jerusalem, August 1997.

\bibitem{CLEO}
CLEO Collaboration, R.~Godang {\it et al.}, 
Phys. Rev. Lett. {\bf 79} (1997) 3829.

\bibitem{PDG}
Particle Data Group, Review of Particle Physics, C.~Caso {\it et al.}
Eur. Phys. J. {\bf C 3} (1998) 1. 

\bibitem{Brodsky}
S.J.~Brodsky, T.~Kinoshita, and H.~Terazawa,
Phys. Rev. {\bf D 4} (1971) 1532.

\bibitem{AAgamma}
G.~Baur and L.G.~Ferreira Filho, Nucl. Phys. {\bf A 518} (1990) 786;
R.N.~Cahn and J.D.~Jackson, Phys. Rev. {\bf D 42} (1990) 3690;
M.~Vidovic, M.~Greiner, C.~Best, and G.~Soff, Phys. Rev. {\bf C 47} 
(1993) 2308;
K.~Hencken, D.~Trautmann, and G.~Baur, Z. Phys. {\bf C 68} (1995) 473.
See also G.~Baur, these proceedings. 

\bibitem{Jackson}
Classical Electrodynamics, J.D.~Jackson (John Wiley \& Sons, 1975).

\bibitem{Glauber}
R.J.~Glauber in Lectures in Theoretical Physics, Eds. W.E.~Brittin and
L.G.~Dunham (Interscience, New York, 1959); C.Y.~Wong, Introduction to
High--Energy Heavy--Ion Collisions (World Scientific, 1994).

\bibitem{nucsize}
Nuclear Sizes and Structure, R.C.~Barrett and D.F.~Jackson
(Oxford University Press, 1977).

\bibitem{Brodsky94}
S.J.~Brodsky {\it et al.},
Phys. Rev. {\bf D 50} (1994) 3134.

\bibitem{Ryskin97}
M.G.~Ryskin, R.G.~Roberts, A.D.~Martin, and E.M.~Levin,
Z.~Phys. {\bf C 76} (1997) 231.

\bibitem{Baur93}
N.~Baron and G.~Baur, Phys. Rev. {\bf C 48} (1993) 1999.

\bibitem{Ellis}
M.~Drees, J.~Ellis, and D.~Zeppenfeld, Phys. Lett. {\bf B 223} (1989) 454.

\bibitem{Prep}
S.~Klein and J.~Nystrand, in preparation.

\bibitem{Crittenden} 
Exclusive Production of Neutral Vector Mesons at the Electron--Proton 
Collider HERA, J.A.~Crittenden
(Springer--Verlag, 1997).

\bibitem{Freund}
P.G.O.~Freund, Nuovo Cimento {\bf A 48} (1967) 541.

\bibitem{Zeus98}
ZEUS Collaboration, J.~Breitweg {\it et al.}, Eur. Phys. J. {\bf C 2} (1998) 
247.

\bibitem{Donnachie}
A.~Donnachie and G.~Shaw, in Electromagnetic Interactions of Hadrons,
Eds. A.~Donnachie and G.~Shaw (Plenum Press, 1978), Vol.~2, p.~169.

\bibitem{starnote}
S.~Klein and J.~Nystrand, STAR Note 347, June 1998, available on the WWW
at http://rsgi01.rhic.bnl.gov/star/starlib/doc/www/sno/ice/sn0347.html.

\bibitem{STARLight}
S.~Klein and E.~Scannapieco, in Proc. Photon '97, Egmond aan Zee, 
The Netherlands, May 10--15, 1997, Eds. A.~Buijs and F.C.~Bern\'e
(World Scientific, 1998), p. 348. \newline hep--ph/9706358.

\bibitem{FRITIOF}
H.~Pi, Comp. Phys. Comm. {\bf 71} (1992) 173. 

\bibitem{VENUS}
K.~Werner, Phys. Rep. {\bf 232} (1993) 87. 

\bibitem{DTUNUC}
R.~Engel, J.~Ranft, and S.~Roesler, Phys. Rev. {\bf D 55} (1997) 6957; 
{\it ibid.} {\bf D 57} (1998) 2889.

\bibitem{HemiCosm}
M.P.~Bringle,  BaBar Note 163 (1994) (unpublished).

\end{thebibliography}
\end{document}